\newcommand{\sla}[1]{/\!\!\!#1}
\documentstyle[12pt,axodraw]{article}
\begin{document}
\begin{titlepage}
\begin{flushright}
UM-TH-96-05\\
March 1996
\end{flushright}
\vskip 3cm
\begin{center}
{\large\bf Gauge Invariance in the Process\\
           $e^+e^-\rightarrow \bar \nu_e e^-W^+
            \rightarrow\bar\nu_e e^-u\bar d$}
\vskip 1cm
{\large Robin G. Stuart}
\vskip 1cm
{\it Randall Laboratory of Physics\\
 University of Michigan\\
 Ann Arbor, MI 48109-1120\\
 USA\\}
\end{center}
\vskip 2.5cm
\begin{abstract}
The process $e^+e^-\rightarrow \bar \nu_e e^-W^+
\rightarrow\bar\nu_e e^-u\bar d$ is considered as
an example of the problems associated with maintaining gauge invariance in
matrix elements involving unstable particles. It is shown how to construct
a matrix element that correctly treats width effects for the intermediate
unstable $W$ boson and that is both $SU(2)_L$
and $U(1)_{\rm e.m.}$ gauge-invariant. $SU(2)_L$ gauge-invariance is
maintained by Laurent expansion in kinematic invariants and $U(1)_{\rm e.m.}$
gauge-invariance is enforced by means of a projection operator under which
the exact matrix element is invariant.
\end{abstract}

\end{titlepage}

\setcounter{footnote}{0}
\setcounter{page}{2}
\setcounter{section}{0}
\newpage

\section{Introduction}

In recent years the difficulty and importance of producing exactly
gauge-invariant amplitudes for processes involving unstable particles
has come to be appreciated. Prior to 1991 all calculations of electroweak
physics at the $Z^0$ resonance, and even the definition of $M_Z$, were
gauge-dependent. A variety of techniques have been applied to try to
remove explicit gauge-dependence. Some of these involve the generation of
gauge-invariant Green's functions, such as self-energies and vertex
corrections \cite{KennedyLynn,KLIS,DegrassiSirlin}.
Such techniques however replace dependence on, say, the
gauge parameter by some arbitrary choice in the definition of the Green's
function. In the case of the so-called `pinch technique' the procedure used
may be rigorously justified by the background field method
\cite{DennWeigDitt1,DennWeigDitt2}
for a particular choice of the gauge parameter, $\xi$. The fact that a
particular choice of gauge is involved indicates that the gauge dependence
has been disguised but not eliminated. The appearance of gauge-dependent
amplitudes is a symptom of a perturbative expansion that does not reflect
the physics of the problem.

The only method to produce genuinely gauge-invariant amplitudes at all
orders in perturbation theory is that of Laurent expansion as proposed
in ref.\cite{Stuart1,Stuart3}. This
expansion should be applied even in the case
of amplitudes generated via the pinch technique. In ref.\cite{Stuart4}
it was pointed out that the expansion was not just a mathematical trick.
The leading resonant term in the expansion represents the finite propagation
and subsequent decay of a physical unstable particle. As such it constitutes
a distinguishable physical process and therefore must be exactly
gauge-invariant. This insight was used to give an answer to the long-standing
problem of how to calculate production cross-sections for unstable particles
and was applied to the process $e^+e^-\rightarrow Z^0Z^0$.

The presence of an unstable particle is indicated by a finite
propagation length separating production and decay vertices. The contributions
responsible for finite propagation are just those that are resonant in the
invariant mass of the decay products. Thus, for example, to calculate the
cross-section $e^+e^-\rightarrow Z^0Z^0$ one calculates the matrix element
for the process
$e^+e^-\rightarrow Z^0Z^0\rightarrow (f_1\bar f_1)(f_2\bar f_2)$
and extracts the part resonant in the invariant masses,
$p_1^2$ and $p_2^2$, of the $(f_1\bar f_1)$ and $(f_2\bar f_2)$ pairs
respectively. This is accomplished by means of a Laurent expansion in each
of the variables $p_1^2$ and $p_2^2$. In the notation of
ref.\cite{BardeenTung} the general matrix element for the process takes
the form
\begin{equation}
{\cal M}(...,p_1,p_2,...)=\sum_i l_i(...,p_1,p_2,...) A_i(t,u,p_1^2,p_2^2)
\label{eq:generalmatelm}
\end{equation}
where the $l_i$ are `standard covariants',
with external wavefunctions attached, that form a basis for the
spinor and Lorentz tensor structure of the matrix element. They
therefore transform as ${\cal M}$ does.
The $A_i$ are Lorentz scalar functions of the independent Lorentz
invariants of the problem and contain no kinematic singularities.
The resonant part is extracted by taking the leading terms
$\hat A_i$ in a Laurent expansion first in $p_1^2$ and then $p_2^2$.

The cross-section formed from the resonant parts
\begin{equation}
\widehat{\cal M}(...,p_1,p_2,...)=
     \sum_i l_i(...,p_1,p_2,...)\hat A_i(t,u,p_1^2,p_2^2)
\label{eq:resonantmatelm}
\end{equation}
or in this case doubly resonant parts, of the matrix element
and then summed over all possible final states constitutes the cross-section
$\sigma(e^+e^-\rightarrow Z^0Z^0)$. If the standard covariants $l_i$ do
form a basis then the scalars $A_i$ will be gauge-invariant and hence
so will the $\hat A_i$.

The external particle wave functions do not feature in the expansion process
hence the kinematics of the problem are unchanged and momentum conservation
is preserved throughout. The final state integrations involve only stable
on-shell particles and so things like complex scattering angles that have
plagued certain other attempts at calculating such cross-sections do not arise.
Problems near `threshold' that have been noted elsewhere
\cite{AeppCuypOlde,AeppOldeWyl} do not appear.
This is as expected since unstable particles do not exhibit sharp thresholds as
do stable ones. Branch cuts corresponding to their production lie off the real
axis and such threshold problems must be calculational artifacts.
There is no ambiguity or flexibility at any stage of the procedure.
Because matrix elements are always evaluated with external stable states they
remain gauge-invariant throughout. As guaranteed by Fredholm theory, there
will be an exact factorization between the final state decay products of the
unstable particle and the rest of the matrix element. Because of this exact
factorization, the matrix element takes the form
\begin{equation}
\sigma(s)=\int_0^s dp_1^2
          \int_0^{(\sqrt{s}-\sqrt{p_1^2})^2} dp_2^2
          \sigma(s;p_1^2,p_2^2)\ \rho(p_1^2)\ \rho(p_2^2),
\label{eq:ZZxsec}
\end{equation}
where $\sigma(s;p_1^2,p_2^2)$ is the cross-section obtained from the
resonant part of the matrix element alone and
\[
\rho(p^2)\approx\frac{1}{\pi}
         .\frac{p^2 (\Gamma_Z/M_Z)}{(p^2-M_Z^2)^2+\Gamma_Z^2 M_Z^2}
          \theta(p_0)\theta(p^2)
\]
is Breit-Wigner-like convolution function.

In ref.\cite{cardinals} the process
$e^+e^-\rightarrow\bar\nu_e e^-W^+\rightarrow\bar\nu_e e^-u\bar d$
is treated as an example of how to include width-effects in matrix
elements involving unstable particles.
Much is made of electromagnetic $U(1)_{\rm e.m.}$
gauge invariance and the possibility that the large mass ratio $s/m^2_e$
present in this process might amplify uncancelled gauge-dependence to
a disastrous level \cite{Berends,Kurihara}.
In particular it is shown that gauge invariance
guarantees that the cross-section behaves as $\sim q^{-2}$ as
$q^2\rightarrow 0$ and not $\sim q^{-4}$ as it might otherwise do.
However the question of how to simultaneously maintain
$SU(2)_L$ and $U(1)_{\rm e.m.}$ gauge invariance in calculations that take
account of $W$~boson width effects is not considered. The approach advocated
is to include a subset of
higher-order corrections to restore $U(1)_{\rm e.m.}$
gauge invariance as was done in ref.s\cite{BaurZeppen,Papadopoulos}.
This procedure is however inconvenient, difficult to apply consistently
and adds considerably to the calculational labour involved.
As noted in ref.\cite{cardinals} the reparation scheme is arbitrary.

It may be commented that the use of next order diagrams to fix problems
at a given order for processes involving unstable particles has been
noted elsewhere \cite{Moriond}. In that case troublesome imaginary parts of
counterterms were observed to be canceled by next order diagrams.

\section{Gauge Invariance for Massless Particles}

When massless gauge particles are present special problems arise.
Suppose we were to write a Green's function $G^\mu$
with a photon leg in terms of some basis of standard covariants, $l_i^\mu$,
and Lorentz scalars, $A_i$, in a form analogous to
eq.(\ref{eq:generalmatelm}) and (\ref{eq:resonantmatelm}),
\begin{equation}
G^\mu(p_1,p_2,...)=\sum_i l_i^\mu(p_1,p_2,...) A_i(p_1^2,p_2^2,...).
\label{eq:greenfunc}
\end{equation}
The $l_i^\mu$, of course, must transform in the same way that $G^\mu$ does.

The gauge condition on Green's functions $q_\mu G^\mu=0$ means that the
$A_i$ are not linearly independent since they satisfy the condition
$\sum_i (q\cdot l_i)A_i=0$ where $q_\mu$ is the momentum of the photon leg.

Gauge-invariant Green's
functions can be obtained using the methods of Bardeen and Tung
\cite{BardeenTung}. Suppose that we have some exact Green's function expressed
in terms of standard covariants and Lorentz scalars as in
eq.(\ref{eq:greenfunc}). Since $q_\mu G^\mu=0$ the Green's function is
invariant under of the action of the operator
\begin{equation}
I^\mu_\nu=g^\mu_\nu-\frac{p^\mu q_\nu}{p\cdot q}
\label{eq:projoper}
\end{equation}
where $p$ is some conveniently chosen momentum.
The spinor-Lorentz tensor structure of the Green's function is spanned by
the set $\{I^\mu_\nu l_i^\nu\}$. That is
\begin{equation}
G^\mu(p_1,p_2,...)=\sum_i \left(I_\nu^\mu l_i^\nu(p_1,p_2,...)\right)
                   A_i(p_1^2,p_2^2,...).
\end{equation}
For an approximate Green's function that does not satisfy the
$U(1)_{\rm e.m.}$ gauge-condition, the operator $I_\nu^\mu$ serves as a
projection operator onto $\{I^\mu_\nu l_i^\nu\}$. From the
$\{I^\mu_\nu l_i^\nu\}$ it is always possible
to construct a new basis that is free of
kinematic singularities that might have occurred when, for example,
$(p\cdot q)=0$.

In general consider a Green's function $G^\mu$ calculated using some
incomplete expansion up to some order. It will consist of two parts
\begin{equation}
G^\mu=G_0^\mu+G_1^\mu
\end{equation}
where $G_0^\mu$ is a consistent gauge-invariant contribution
correct to given order of the calculation. $G_1^\mu$ is a spurious
higher-order gauge-dependent correction. Because $G_0^\mu$
satisfies the gauge condition, $q_\mu G_0^\mu=0$, it is invariant
under the action of the operator $I_\mu^\nu$.
Thus at a given order we may make the replacement
\[
G^\mu\rightarrow I^\mu_\nu G^\nu
             =G_0^\mu+I^\mu_\nu G_1^\nu.
\]
Only $G_1^\mu$ is affected by the projection operator but
since it is of higher order this is of no concern. The important point is
that it now satisfies the gauge condition $q_\mu(I^\mu_\nu G_1^\nu)=0$
and is prevented from causing too strong a numerical blow up as
$q^2\rightarrow0$.

\section{The Process
         $e^+e^-\rightarrow\bar\nu_e e^-W^+\rightarrow\bar\nu_e e^-u\bar d$}

Consider the process
$e^+e^-\rightarrow\bar\nu_e e^-W^+\rightarrow\bar\nu_e e^-u\bar d$.
Some of the tree level diagrams contributing to this process are shown in
Fig.1. In the kinematic regions where the process is allowed
(but $W^+W^-$ production is forbidden) it is expected
to be dominant because of the presence of a resonant
$W$ propagator. We will mainly concern ourselves with the region with
$q^2$ small but the methods used apply in any kinematic region and will
yield results that are exactly $SU(2)_L$ and $U(1)_{\rm e.m.}$
gauge-invariant order by order.

Let $q$ be the 4-momentum of the virtual photon and $p_+$ be that of
the $W^+$. The momenta of the incoming electron and positron are $p_1$ and
and $k_1$ respectively and $p_2$ and $k_2$ are those of the outgoing
electron and $\bar\nu_e$. Hence
\begin{equation}
p_+ = p_u+p_d,\ \ \ \ p_- = k_1-k_2,\ \ \ \ q = p_1-p_2.\\
\end{equation}
$M_W$ is the $W$ mass, $e$ and $g$ are the electromagnetic and $SU(2)_L$
weak coupling constants respectively.
$Q_i$ is the electric charge of particle $i$.
As usual $\gamma_L=\frac{1}{2}(1-\gamma_5)$
is the left-handed helicity operator.


\begin{figure}[htb]
\begin{center}
\begin{picture}(90,100)(0,0)
\ArrowLine(10,90)(35,80)
\ArrowLine(35,80)(80,90)
\ArrowLine(80,10)(35,20)
\ArrowLine(35,20)(10,10)
\ArrowLine(85,35)(70,50)
\ArrowLine(70,50)(85,65)
\Photon(35,80)(40,50){2}{4}
\Photon(40,50)(35,20){2}{4}
\Photon(40,50)(70,50){2}{4}
\Vertex(35,80){1.2}
\Vertex(35,20){1.2}
\Vertex(40,50){1.2}
\Vertex(70,50){1.2}
\put(08,90){\makebox(0,0)[r]{$e^-$}}
\put(08,10){\makebox(0,0)[r]{$e^+$}}
\put(82,90){\makebox(0,0)[l]{$e^-$}}
\put(82,10){\makebox(0,0)[l]{$\bar{\nu}_e$}}
\put(87,65){\makebox(0,0)[l]{$u$}}
\put(87,35){\makebox(0,0)[l]{$\bar{d}$}}
\put(34,65){\makebox(0,0)[r]{$\gamma$}}
\put(35,35){\makebox(0,0)[r]{$W$}}
\put(57,55){\makebox(0,0)[b]{$W^+$}}
\end{picture}
\qquad
\begin{picture}(90,90)(0,-10)
\ArrowLine(10,70)(35,60)
\ArrowLine(35,60)(80,70)
\ArrowLine(80, 0)(55,18)
\ArrowLine(55,18)(35,20)
\ArrowLine(35,20)(10,10)
\ArrowLine(85,25)(70,35)
\ArrowLine(70,35)(85,50)
\Photon(35,60)(35,20){2}{5}
\Photon(55,18)(70,35){2}{3}
\Vertex(35,60){1.2}
\Vertex(35,20){1.2}
\Vertex(55,18){1.2}
\Vertex(70,35){1.2}
\put(08,70){\makebox(0,0)[r]{$e^-$}}
\put(08,10){\makebox(0,0)[r]{$e^+$}}
\put(82,70){\makebox(0,0)[l]{$e^-$}}
\put(82, 0){\makebox(0,0)[l]{$\bar{\nu}_e$}}
\put(87,50){\makebox(0,0)[l]{$u$}}
\put(87,25){\makebox(0,0)[l]{$\bar{d}$}}
\put(30,40){\makebox(0,0)[r]{$\gamma$}}
\put(70,30){\makebox(0,0)[br]{$W^+$}}
\end{picture}
\\
\end{center}
\caption[]{Tree-level diagrams contributing to the resonant part of the process
           $e^+e^-\to\bar\nu_e e^-W^+\to\bar\nu_e e^-u\bar d$.}
\end{figure}
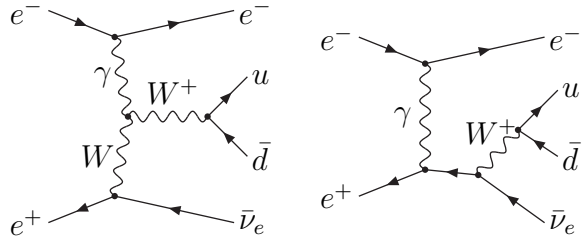

Using the same notation as in the introduction the exact matrix element takes
the form
\begin{equation}
{\cal M}=\sum_i l_i(q,p_+,...) A_i(q^2,p_+^2,...)
\end{equation}
This matrix element may be divided into a part that has a simple pole at
$q^2=0$ and a part that has no pole
\begin{equation}
{\cal M}=\sum_i l_i(q,p_+,...) \frac{R_i(0,p_+^2,...)}{q^2}
        +\sum_i l_i(q,p_+,...) \frac{R_i(q^2,p_+^2,...)
                                      -R_i(0,p_+^2,...)}{q^2}.
\label{eq:qsqexp}
\end{equation}
These terms must be separately $SU(2)_L$ gauge-invariant because their
differing pole structures. The first term
receives contributions from diagrams, such as those of Fig.1, in which a
photon is connected to the external electron current. The second term receives
contributions from these as well many other topologies. Examples are the
diagrams of Fig.1 in which the photon is replaced by a $Z^0$ boson.

We will be interested in the small $q^2$ region and will thus concentrate
on the first term in eq.(\ref{eq:qsqexp}) and write
\begin{equation}
{\cal M}=\sum_i J_\mu^\gamma (l_i^\prime)^\mu(q,p_+,...)
                             R_i^\prime(0,p_+^2,...)
\label{eq:photonmat}
\end{equation}
where
\begin{equation}
J^\gamma_\mu=\frac{V_\gamma(0)}
                  {[1-\Pi_{\gamma\gamma}^\prime(0)]^{\frac{1}{2}}}.
             \frac{\bar u(p_2)\gamma_\mu u(p_1)}{q^2}.
\end{equation}
The $(l_i^\prime)^\mu(q,p_+,...)$ are standard covariants with attached
external wavefunctions and $V_\gamma(q^2)$ is the $\gamma e^+e^-$
vertex form-factor. $\Pi_{\gamma\gamma}^\prime(q^2)$ is the derivative of
the photon self-energy with respect to $q^2$. The factorization of the
matrix element in this way is guaranteed by Fredholm theory.
Since ${\cal M}$ is $U(1)_{\rm e.m.}$ gauge
invariant it is possible to find a set $\{(l_i^\prime)^\mu(q,p_+,...)\}$
that is free of kinematic singularities and that
satisfy $q_\mu(l_i^\prime)^\mu(q,p_+,...)=0$ \cite{BardeenTung}.

>From the part of the matrix element (\ref{eq:photonmat})
we wish to extract the dominant resonant
part of the matrix element that describes the production of a physical $W^+$
and its subsequent decay into the $\bar u d$ final state. As with the process
$e^+e^-\rightarrow Z^0Z^0$ this is the leading term in a
Laurent expansion of $R_i(0,p_+^2,...)$ in $p_+^2$, the invariant mass of the
$W^+$ decay products. The result takes the form
\begin{equation}
{\cal M}=\sum_i J_\mu^\gamma J_\alpha^W
                (l_i^{\prime\prime})^{\mu\alpha}(q,p_+,...)
                R_i^{\prime\prime}(0,s_W,...)
\label{eq:exactexp}
\end{equation}
where
\begin{equation}
J^W_\alpha=\frac{V_W(s_W)}
                {[1-\Pi_{WW}^\prime(s_W)]^{\frac{1}{2}}}.
           \frac{\bar{u}(p_u)\gamma_\alpha\gamma_L v(p_d)}
                      {p_+^2-s_W}
\end{equation}
Here $\Pi_{WW}^\prime(q^2)$ is the derivative of the $W$ self energy with
respect to $q^2$ and
$s_W$ the solution of $p_+^2-M_W^2-\Pi_{WW}(p_+^2)=0$. $V_W(p_+^2)$
is $W^+u\bar d$ vertex form factor. The result (\ref{eq:exactexp}) is
exactly $SU(2)_L$ and $U(1)_{\rm e.m.}$ gauge-invariant.

Since the $(l_i^{\prime\prime})^{\mu\alpha}(q,p_+,...)$ are
$U(1)_{\rm e.m.}$ gauge-invariant they are invariant under the action
of projection operators of the form given in eq.(\ref{eq:projoper}).

In finite order calculations, because of the non-perturbative nature of
the $W^+$ resonance, $U(1)_{\rm e.m.}$ gauge-invariance may not be
satisfied. A gauge-invariant result may
be obtained by inserting a projection operator, $I_\nu^\mu$, as defined
in eq.(\ref{eq:projoper}). This has the the effect
of discarding  spurious gauge-dependent terms that must ultimately cancel
in the final result thus avoiding the need to calculate higher-order
diagrams as was done in ref.s\cite{cardinals,BaurZeppen}.

With these general considerations in mind we can now calculate the
cross-section for the process
$\sigma(e^+e^-\rightarrow \bar\nu_e e^-W^+
              \rightarrow \bar\nu_e e^-u\bar d)$
in leading order and with the $W$ width effects correctly taken into account.
In words this is the production cross-section for $e^-\bar\nu_e$ and a
physical $W^+$ that subsequently decays into a $u\bar d$ pair.

As discussed above the contribution relevant as $q^2\rightarrow 0$ takes the
form
\begin{equation}
{\cal M}={\cal M}^{\mu\alpha}J^\gamma_\mu J^W_\alpha.
\label{eq:matelm}
\end{equation}
in which
\begin{equation}
{\cal M}^{\mu\alpha}=\sum_{i=1}^2{\cal M}_i^{\mu\alpha}.
\end{equation}
The ${\cal M}_i^{\mu\alpha}$ correspond to the two diagrams of Fig.1,
\begin{eqnarray}
{\cal M}_1^{\mu\alpha}&=&ie\frac{g}{\sqrt{2}}
  Q_W\frac{1}{p_-^2-M_W^2}
  V^{\alpha\beta\mu}(p_+,-p_-,-q)
  \bar{v}(k_1)\gamma_\beta \gamma_L v(k_2),\label{eq:M1}\\
{\cal M}_2^{\mu\alpha}&=&ie\frac{g}{\sqrt{2}}
  Q_e\bar{v}(k_1)\gamma^\mu{\sla{k}_1+\sla{q}\over(k_1+q)^2}
     \gamma^\alpha \gamma_L v(k_2).\label{eq:M2}\\
\end{eqnarray}
and
\begin{equation}
V^{\mu_1\mu_2\mu_3}(p_1,p_2,p_3) = (p_1-p_2)^{\mu_3}g^{\mu_1\mu_2} +
(p_2-p_3)^{\mu_1}g^{\mu_2\mu_3} + (p_3-p_1)^{\mu_2}g^{\mu_3\mu_1}.
\end{equation}
The projection operator $I_\nu^\mu$ of eq.(\ref{eq:projoper})
may be inserted into eq.(\ref{eq:matelm})
in order to guarantee $U(1)_{\rm e.m.}$ gauge invariance of the Green's
function, ${\cal M}^{\mu\alpha}$. The matrix element then becomes
\begin{equation}
{\cal M}={\cal M}^{\mu\alpha}I_\mu^\lambda J^\gamma_\lambda J^W_\alpha.
\end{equation}

The matrix element squared and averaged over spins then takes the form
\begin{equation}
\left\langle|{\cal M}|^2\right\rangle=
     \frac{1}{4}\sum_{\mbox{\tiny $e^+e^-$ spins}}
                {\cal M}^{\mu\alpha} I_\mu^\lambda
                J^\gamma_\lambda g_{\alpha\beta}\rho(p_+^2)
\overline{{\cal M}^{\nu\beta}I_\nu^\rho J^\gamma_\rho}
\end{equation}
with
\begin{eqnarray}
g_{\alpha\beta}\rho(p_+^2)
&=&\sum_{\mbox{\tiny $u\bar d$ spins}}\int\frac{d^3p_u}{(2\pi)^3 2p^0_u}
                             \frac{d^3p_d}{(2\pi)^3 2p^0_d}
J^W_\alpha\overline{J^W_\beta}
(2\pi)^4\delta^4(p_u+p_d-p_+)\theta(p_+^0)\theta(p_+^2)\nonumber\\
&=&\frac{g^2}{48\pi^2}.
   \frac{p_+^2}{|p_+^2-s_W|^2} g_{\alpha\beta}\theta(p_+^0)\theta(p_+^2)
\end{eqnarray}
at leading order.

In forming the total production cross-section for a physical $W^+$'s that
subsequently decays into $u\bar d$ we must sum over the $u\bar d$ final
states. Hence
\begin{equation}
\sigma(e^+e^-\rightarrow\bar\nu_e e^-W^+\rightarrow\bar\nu_e e^-u\bar d)
      =N_c\int dp^2\sigma(p^2)\rho(p^2).
\label{eq:crosssection}
\end{equation}
The integral is over the full range allowed by the cut on the
scattering angle, $\theta$. $N_c$ is the number of QCD colours and
\begin{eqnarray}
\sigma(p^2)&=&\frac{1}{2s}\int\frac{d^3p_2}{(2\pi)^3 2p^0_2}
                \frac{d^3k_2}{(2\pi)^3 2k^0_2}
                \frac{d^3p_+}{(2\pi)^3 2p^0_+}
                \left\langle|{\cal M}_0|^2\right\rangle\nonumber\\
& &\ \ \ \ \ \times(2\pi)^4\delta^4(p_1+k_1-p_2-k_2-p_+)
 \end{eqnarray}
in which
\begin{equation}
\left\langle|{\cal M}_0|^2\right\rangle=
     \frac{1}{4}\sum_{\mbox{\tiny $e^+e^-$ spins}}
                {\cal M}^{\mu\alpha}I_\mu^\lambda J^\gamma_\lambda
 \overline{{\cal M}^{\nu\alpha}I_\nu^\rho J^\gamma_\rho}
\end{equation}
and the integral is evaluated under the constraint that $p_+^2=p^2$.
Because ${\cal M}^{\mu\alpha}I_\mu^\lambda$ is $U(1)_{\rm e.m.}$
gauge-invariant, $\left\langle|{\cal M}_0|^2\right\rangle\sim q^{-2}$
only as $q^2\rightarrow 0$.

\section{Acknowledgments}
The author wishes to thank D.\ Williams for many enlightening discussions
and the authors of ref.\cite{cardinals} for permitting use of their figures
of Feynman diagrams.

\end{document}